\documentclass[useAMS,usenatbib,usegraphicx]{mn2e}
\usepackage{natbib}
\usepackage{amsmath}
\usepackage{amssymb}
\newcommand{\bo}{B\"O02}
\newcommand{\z}{Z06}
\newcommand{\arcsecsq}{$~\textrm{arcsec}^{-2}$}
\title[Haro 11: is there a red excess?]{The faint outskirts of the blue compact galaxy Haro 11: is there a red excess?}
\author[Micheva et al.]{Genoveva Micheva$^{1}$\thanks{E-mail: genoveva@astro.su.se (GM)}, Erik Zackrisson$^{2}$\thanks{E-mail: ez@astro.su.se (EZ)}, G\"oran \"Ostlin$^{2}$\thanks{E-mail: ostlin@astro.su.se (G\"O)}, Nils Bergvall$^{3}$\thanks{E-mail: nisse@fysast.uu.se} and Tapio Pursimo$^{4}$\thanks{E-mail: tpursimo@not.iac.es}\\
$^{1}$Stockholm Observatory, Department of Astronomy, Stockholm University, 106 91 Stockholm, Sweden\\
$^{2}$Oskar Klein Centre for Cosmoparticle Physics, Department of Astronomy, Stockholm University, 106 91 Stockholm, Sweden\\
$^{3}$Division of Astronomy \& Space Physics, Uppsala university, 751 20 Uppsala, Sweden\\
$^{4}$Nordic Optical Telescope, Apdo 474, 38700 Santa Cruz de La Palma, Spain}
\begin{document}

\date{Accepted .... Received ...; in original form ...}

\pagerange{\pageref{firstpage}--\pageref{lastpage}} \pubyear{2008}

\maketitle

\label{firstpage}

\begin{abstract}

\noindent Previous studies of the low surface brightness host of the blue compact galaxy (BCG) Haro 11 have suggested an abnormally red color of $V-K=4.2\pm0.8$ for the host galaxy. This color is inconsistent with any normal stellar population over a wide range of stellar metallicities ($Z=0.001$--$0.02$). Similar though less extreme host colors have been measured for other BCGs and may be reconciled with population synthesis models, provided that the stellar metallicity of the host is higher than that of the ionized gas in the central starburst. We present the deepest V and K band observations to date of Haro 11 and derive a new $V-K$ color for the host galaxy. Our new data suggest a far less extreme colour of $V-K=2.3\pm0.2$, which is perfectly consistent with the expectations for an old host galaxy with the same metallicty as that derived from nebular emission lines in the star-forming center.
\end{abstract}


\section{Introduction}

\noindent BCGs are gas-rich low-luminosity galaxies of particular interest since many of them have very high star formation rates (SFR) and low chemical abundances~\citep{Kunth_Ostlin_2000}. They are thus reminiscent of high-redshift starbursting young galaxies. This, together with their close proximity to us, makes BCGs useful test objects, suitable for gaining insight into galaxy formation and intense star formation (SF). Aside from the bright central starburst, deep photometric optical and near-infrared (NIR) studies~\citep[e.g.][]{Loose_Thuan_1986, Doublier_etal_1997, Papaderos_etal1_1996, Cairos_etal1_2001,Bergvall_Ostlin_2002,Amorin_etal_2007,Amorin_etal_2009,Noeske_etal_2005} have revealed the presence of another component of low surface brightness (LSB). The LSB component hosts the central starburst and is therefore referred to as a LSB host galaxy. The host is only visible close to the outskirts of BCGs since at small radial distances the intensity from the central starburst outshines the contribution from the LSB component. Many photometric studies have concentrated on reaching faint levels in an attempt to isolate the LSB component and characterize it. Some have suggested that the LSB component has unusual properties such as an extreme red excess in the optical/NIR colors.~\citet[][ hereafter \bo]{Bergvall_Ostlin_2002} and~\cite{Bergvall_etal_2005} present a sample of 10 BCGs with progressively redder colors towards the outskirts of the host. \bo~and~\citet[][ hereafter \z]{Zackrisson_etal_2006} argue that these colors are much too red to be reconciled with a normal stellar population.

\noindent Extended faint structures of very red colors have also been detected in the outskirts of disk galaxies and, similarly to BCGs, these structures display colors much too red to be due to stellar populations with reasonable metallicities and normal initial mass functions (IMFs). In 2004,~\citeauthor*{Zibetti_etal_2004} reported a faint structure with a notable red excess around the stacked disk of 1047 edge-on spirals from the Sloan Digital Sky Survey (SDSS). In the same year,~\citeauthor*{Zibetti_Ferguson_2004} reported a red structure around a disk galaxy in the Hubble Ultra Deep Field (HUDF). Similarly,~\citeauthor*{BergvallZackrissonCaldwell2009}~\citeyearpar{BergvallZackrissonCaldwell2009} find a red excess in the outskirts of disk galaxies by stacking $1510$ edge-on LSB galaxies from the SDSS.

\noindent These detections have been collectively dubbed ``red halos'' and a number of possible explanations have been proposed. Some detections have been falsified and ascribed to instrumental effects. For instance,~\citet{deJong_2008} demonstrated that the HUDF red halo is likely due to reddening by the far wings of the point spread function (PSF). Most detections, however, have persisted and even given rise to exotic scenarios.~\bo~proposed that the extreme colors seen in their sample could be explained by a metal rich ($> Z_{\odot}$) halo stellar population, in contrast to the low nebular chemical abundances. \z~argued that the only stellar population which explains the published colors of the halos of both disk galaxies and BCGs is one with an extremely bottom-heavy ($dN/dM \propto M^{-\alpha}$ with $\alpha=4.50$) IMF. If proven to be a common component of galaxies of various Hubble types, this would imply that faint low-mass stars are so numerous in these galaxies that they may significantly contribute to the baryons still missing from inventories in the low-redshift Universe. Another effect which may produce red halos around both disk galaxies and BCGs is that the surface brightness profiles used to derive the colors have been oversubtracted due to failure to account for extinction by the extragalactic background light~\citep*{Zackrisson_etal_2009a}.

\noindent If one is willing to trade off a common explanation for all red halo detections with individual explanations applicable to certain galaxies and not to others then most of the red halos of the \bo~and ~\citet{Bergvall_etal_2005} BCGs sample can be modeled with a Salpeter IMF, with the caveat that the required stellar metallicity is in some cases much higher than what is observed from the gas in the central starburst~\citep{Zackrisson_etal_2009b}. Among the sample, one of the biggest, most massive and most luminous BCGs is Haro 11, with a very high SFR of 18-20 $M_{\odot}$yr${}^{-1}$ and the reddest color of $V-K=4.2\pm0.8$, which cannot be reconciled with any normal stellar population.

\noindent Throughout this paper we assume a Hubble parameter of $H_0=75~km~s^{-1}~Mpc^{-1}$ and a distance to Haro 11 of $82~Mpc$, with a redshift of $z=0.020598$. In $\S$~\ref{data} we present the deepest observations to date for Haro 11 in the V and K passbands and a new and more accurate measurement of the $V-K$ color. We are pushing the limits of surface photometry down to isophotes 10000 times fainter than the brightness of the night sky in K and we therefore examine the reliability of the obtained profiles in $\S$~\ref{sbsection} and $\S$~\ref{reliability}. Our results for the $V-K$ color of Haro 11 are presented in $\S$~\ref{results}. Discussion and summary can be found in $\S$~\ref{discuss} and $\S$~\ref{summary}, respectively.

\begin{table*}
 \centering
 \begin{minipage}{140mm}
  \begin{tabular}{@{}lccccccccc@{}}
  \hline
  &\multicolumn{2}{c}{Old data}&\multicolumn{2}{c}{New data}\\
  &V&K${}^\prime$&V&Ks\\
  Year&1984&1993&2008&2005&\\
  Exposure Time&15 min&66 min&40 min&60 min&\\
  Instrument&ESO 2.2m &ESO 2.2m IRAC2&NOT 2.56m MOSCA&ESO NTT 3.58m SOFI&\\
  $\prime\prime$/pixel&0.362&0.475&0.217&0.288&\\
  FoV($\prime\prime$)&178$\times$178&136$\times$136&462$\times$462&295$\times$295&\\
\hline
\end{tabular}
\caption{Observational data summary for Haro 11 (ESO350--IG038). The effective exposure time for new V (new K) data is 3.3 (2.4) times longer than for the old V (old K) data. Haro 11 has an apparent angular diameter of $\sim1$ arcmin out to a surface brightness of $\mu_V=26.5$ mag\arcsecsq. All observations are in dithering mode.}\protect\label{datatbl}
\end{minipage}
\end{table*}


\section[]{Data}\protect\label{data}

\noindent Table~\ref{datatbl} summarizes the available data used in this paper and compares with the data published by~\cite{Bergvall_Ostlin_2002}. The V band data was obtained in 2008 at the Nordic Optical Telescope (NOT) with the MOSaic CAmera (MOSCA), which has a scale of 0.217 arcsec/pixel. The K band data was obtained in 2005 at the ESO NTT 3.58m telescope with Son OF Isaac (SOFI), which has a scale of 0.288 arcsec/pixel. The raw frame size in the optical (MOSCA) is 7.7 arcmin, in the NIR (SOFI) 4.9 arcmin. The NIR images are dithered with large offsets. Due to the large field of view of the instrument and the comparatively much smaller target size, this is the optimal observational strategy for this object.

\subsection{Reductions}

\noindent The data were reduced in the following way. The raw images in both bands were cleaned from bad pixels, bias subtracted in the V band, pair subtracted in the K band using the sky median as a scaling factor, trimmed, flatfielded with an illumination-corrected normalized masterflat, and sky subtracted with a flat surface interpolated with a first order polynomial. In the V band the final stacked image was calibrated with secondary standard stars against the published data by \bo. In the K band each sky-subtracted frame was calibrated against 2MASS stars located in the frame, whereby a minimum of five 2MASS stars per frame were used to obtain its zeropoint. The calibrated K band data was therefore independent of the photometric conditions, even though an examination of the temporal zeropoint variation throughout the observations revealed a scatter of only $\sim0.049$ magnitudes. In both bands, images were registered using Pyraf GEOMAP/GEOTRAN. When shifting and rotating data we interpolated with a 5th order polynomial in order to avoid the Moir\'e effect caused by the default bilinear interpolation. The images were median and average combined without the use of a rejection algorithm. Comparison between the median and average stacked images revealed an insignificant difference in the aperture photometry of point sources on the order of $\sim0.02$ magnitudes. Median stacking has the added advantage of filtering out any lingering cosmic rays and bad pixels, therefore the analysis in this paper was carried out on the median stacked images in both bands. Sky subtraction  was performed on the final stacked image in each band in order to remove any residual sky. When performing sky subtraction we neglected the effects of dust extinction of the extragalactic background light in the outskirts of the host galaxy~\citep*{Zackrisson_etal_2009a} and assumed that the sky over the target can be interpolated from sky regions around the target. The final reduced images in V and K have an effective field of view smaller than the field of view of the detectors in both bands, which is a consequence of the median stacking of dithered images.

\subsection{Sky estimation}
\noindent The shape of a radial color profile is sensitive to the slope of the two surface brightness profiles from which it has been derived. In turn, the slope of the surface brightness profiles is sensitive to any systematic errors in the sky subtraction, especially at large radial distances and very faint isophotal levels. In particular, over- or undersubtracting the sky background would lead to spurious color gradients. It is therefore important to have control over all aspects of the sky background estimation. We briefly outline the major steps involved in this procedure. 

\noindent For each frame the sky is approximated by fitting a flat surface to regions free from sources (``sky regions'') with the PyMidas~\citep{Hook_etal_2006} procedure FitFlat. The placement and size of these sky regions are very important - one aims to have suitably large sky regions uniformly distributed over the frame in order for the interpolated sky surface to be a well-sampled approximation of the sky. The size of the sky regions together with the order of the interpolating function determine whether the interpolated surface will mimic the large scale or the small scale structure of the sky. It is difficult to apply a reasonable model to small scale sky structures. We have chosen to remove only the large scale structures, but include the remaining small scale sky fluctuations in the error analysis ($\S$~\ref{sbsection}).

\noindent In order to model the large scale sky structure of a frame, the area of each sky region is not allowed to drop below $600$ pixels${}^2$ and the interpolating function is a polynomial of first order in both $x$ and $y$. The position and size of the sky regions are determined automatically by the following algorithm. All positive and negative sources are masked out, giving a binary mask for each image. Initially, a very coarse fixed grid is placed over the mask image. Defining sky regions over a frame thus translates to iterating over gridboxes and, at each iteration, determining the maximum size of the sky region that can be placed inside each gridbox. The size of the sky region depends on the presence of sources. If sources are absent, the sky region takes the shape and size of the entire gridbox. If sources are present, a box of smaller size is generated and tested for sources. Smaller boxes are first generated by splitting the gridbox at an iteratively moving point along one axis while keeping the size of the box along the other axis at maximum.  These boxes are thus always anchored at two grid vertices along the same axis. If all of the resulting boxes test positive for sources new boxes are generated by moving a strip of fixed width along one axis and maximum height in the other axis. These boxes may or may not be anchored to any vertices but always touch two opposing gridbox axes. If this too fails, new boxes are generated by varying their size along both axes. These boxes are not adjacent to vertices or to gridbox axes and instead ``float'' in the gridbox interior. If still no boxes free from sources are found the gridbox is simply split in four quadrants along the middle of both its axes, resulting in four boxes of a quarter size. If these quarter boxes also test positive for sources, then no sky region can be defined over this gridbox and the algorithm continues to the next gridbox. 

\noindent After going through the entire grid each quadrant of the frame is checked for the number of obtained boxes. If the number of sky boxes in any quadrant is less than 20, the grid is refined and iteration starts over. The algorithm is such that the exact position, shape and size of the present sources is irrelevant. The imposed limit on the minimum number of sky boxes in each frame quadrant ensures that the tiepoints for the interpolated surface are well distributed to cover the entire frame. 

\noindent We have also carried out tests to evaluate the performance of the automated sky box placement algorithm. These consisted of varying the box shape, box size, and the number of boxes per quadrant, and comparing the effect that would have on the PyMidas FitFlat sky estimation. We conclude that the algorithm is stable and reliable in all of our test cases with the caveat that spatially non-linear low frequency sky variations are intrinsically impossible to perfectly model with a first order polynomial regardless of the choice in placement and size of sky boxes. While we have used PyMidas FitFlat for the sky subtraction itself, a test with IRAF IMSURFIT gives nearly identical results, provided that the same sky box distribution is used in both cases.

\begin{figure*}
  \centering
  \vspace{8.5cm}
  \includegraphics{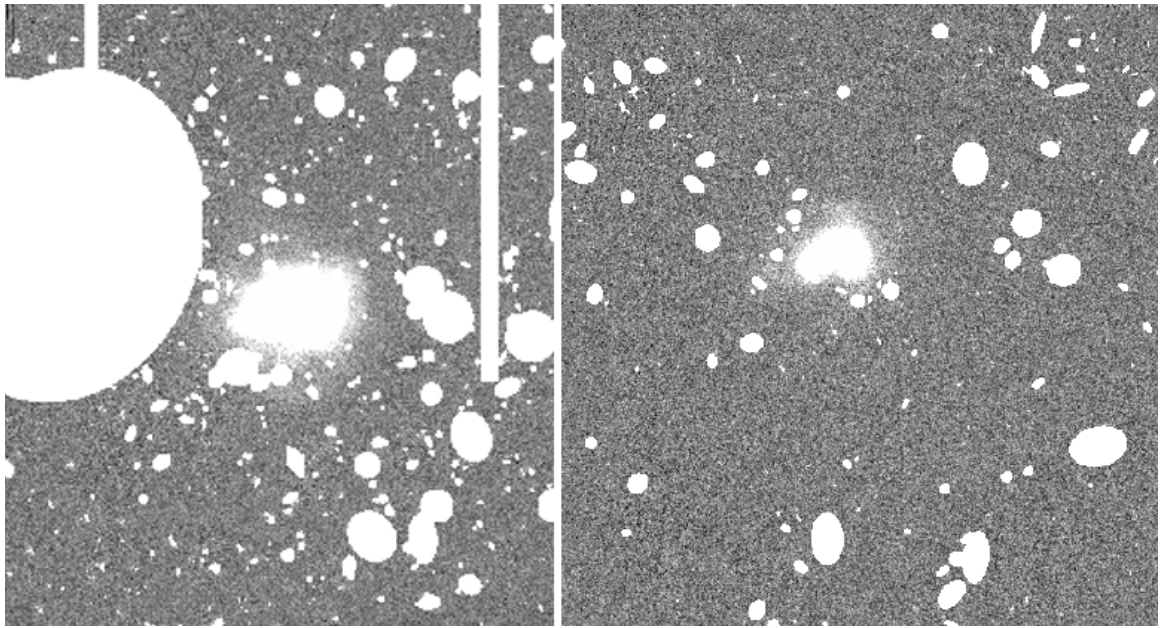}
\caption{New V and K band data for Haro 11. Left panel: NOT MOSCA V band with scale 0.217\arcsec/pixel, 40 minute exposure, 1 chip. The image size is 2.7\arcmin$\times~$2.3\arcmin. Right panel: ESO NTT SOFI K band with scale 0.288\arcsec/pixel, 1 hour exposure. The image size is 2.3\arcmin$\times~$2.2\arcmin. For display purposes the K band image (right) has been scaled to match the size of the V band image (Note that the images have different center coodinates). Bad pixels, sources and cosmic rays in both images are masked out. The larger size of the images significantly increases the area available for the estimation of the sky background. North is up, east is to the left. The large elliptical mask in the V band image is due to a bright blue star. 
}\protect\label{VK}
\end{figure*}

\begin{figure*}
  \centering
  \vspace{8.5cm}
\includegraphics{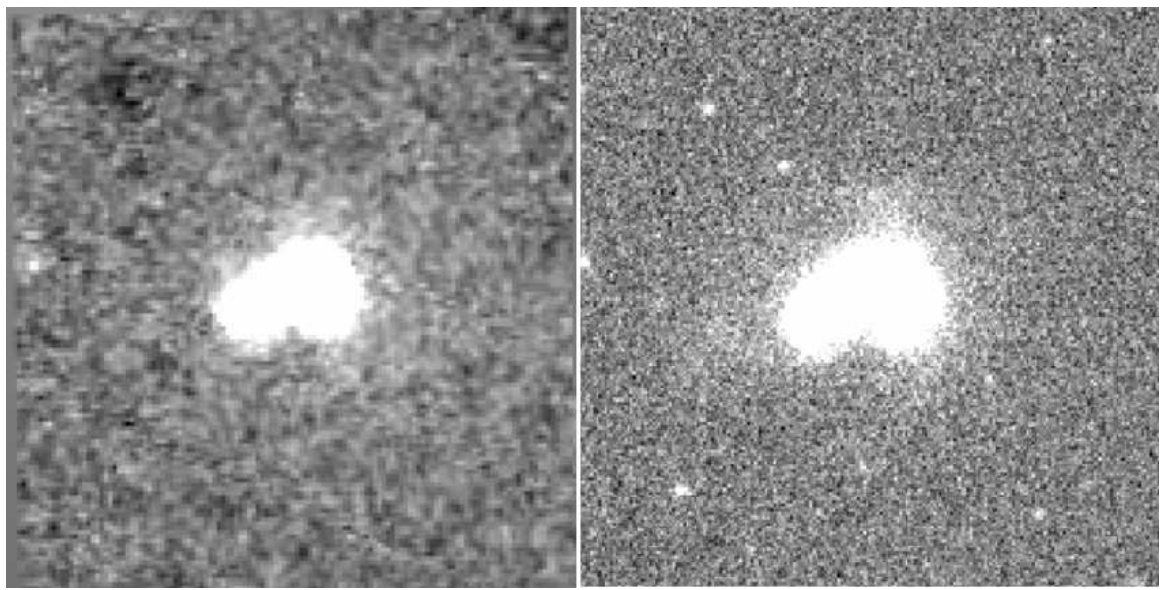}
  \caption{Comparison of the K band data for Haro 11. Left panel: the data used in the analysis of~\protect\cite{Bergvall_Ostlin_2002}. The image size is $\sim$1\arcmin$\times$1\arcmin~with a scale of 0.475 arcsec/pixel. Right panel: the central $\sim$1\arcmin$\times$1\arcmin~ of the new K band data with a scale of 0.288 arcsec/pixel. North is up, east is to the left. The new image is of better quality, with higher resolution and a flatter background.}\protect\label{Kband}
\end{figure*} 

\begin{figure}
  \centering
\includegraphics[width=8.5cm]{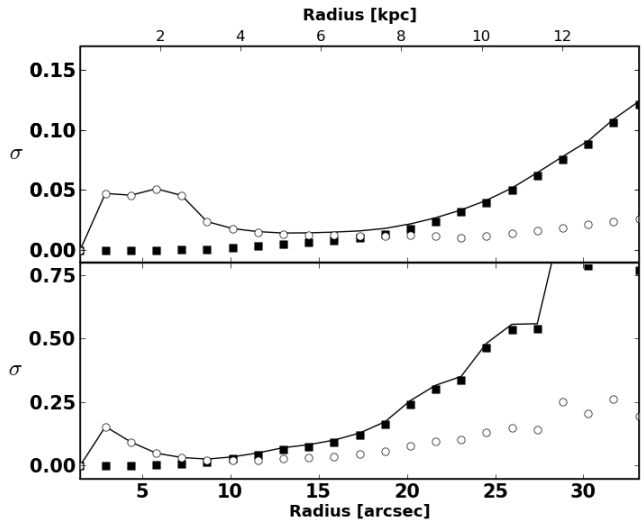}
  \caption[]{\textbf{Radial error composition}. Uncertainties in magnitudes vs. distance in arcseconds from the center of Haro 11 in the V band (upper panel) and the K band (lower panel). The uncertainty in the sky, $\sigma_{\textrm{sky}}$, is marked with filled squares; the standard deviation of the mean, $\sigma_{\textrm{SDOM}}$, with open circles; the total error, $\sigma_{\textrm{tot}}$, with an uninterrupted line. The large contribution of $\sigma_{\textrm{SDOM}}$ at small radii is due to resolved physical structure present at the center of Haro 11. For both bands $\sigma_{\textrm{sky}}$ dominates the total error at large radii.
  }\protect\label{errplot}
\end{figure}
\begin{table}
 \centering
   \begin{tabular}{|l|l|}
  \hline
  $\sigma_{\textrm{sky,V}}$ & 0.016\\
  $\sigma_{\textrm{sky,K}}$ & 0.185\\
  $\sigma_{\textrm{sky,V-K}}$ & 0.185\\
 \hline
  $\sigma_\textrm{SDOM,V}$&0.008\\
  $\sigma_\textrm{SDOM,K}$&0.023\\
  $\sigma_\textrm{SDOM,V-K}$&0.024\\
 \hline
 $\sigma_{\textrm{tot,V}}$ & 0.018\\
 $\sigma_{\textrm{tot,K}}$ & 0.186\\
  $\sigma_{\textrm{tot,V-K}}$ & 0.187\\
  \hline
\end{tabular}
\caption{Error composition of the host defined over the radial range 6-10~kpc. All error values are in magnitudes. \textbf{$\sigma_\textrm{sky}$} is the uncertainty in the sky background, estimated from the standard deviation of the mean values of 275 sky regions in K and 280 sky regions in V. The range of sizes of the sky regions is $(12\times13,16\times21,15\times27,21\times21,27\times27)$~pixels. \textbf{$\sigma_\textrm{SDOM}$} is the standard deviation of the mean of the flux inside the elliptic ring between 6 and 10 kpc. \textbf{$\sigma_{\textrm{tot}}$ } is the composite error, with $\sigma_\textrm{sky}$~and $\sigma_\textrm{SDOM}$ added in quadrature. }\protect\label{errtbl}
\end{table}

\subsection{Surface brightness profiles}
\begin{figure*}
  \centering
  \vspace{7cm}
  \includegraphics{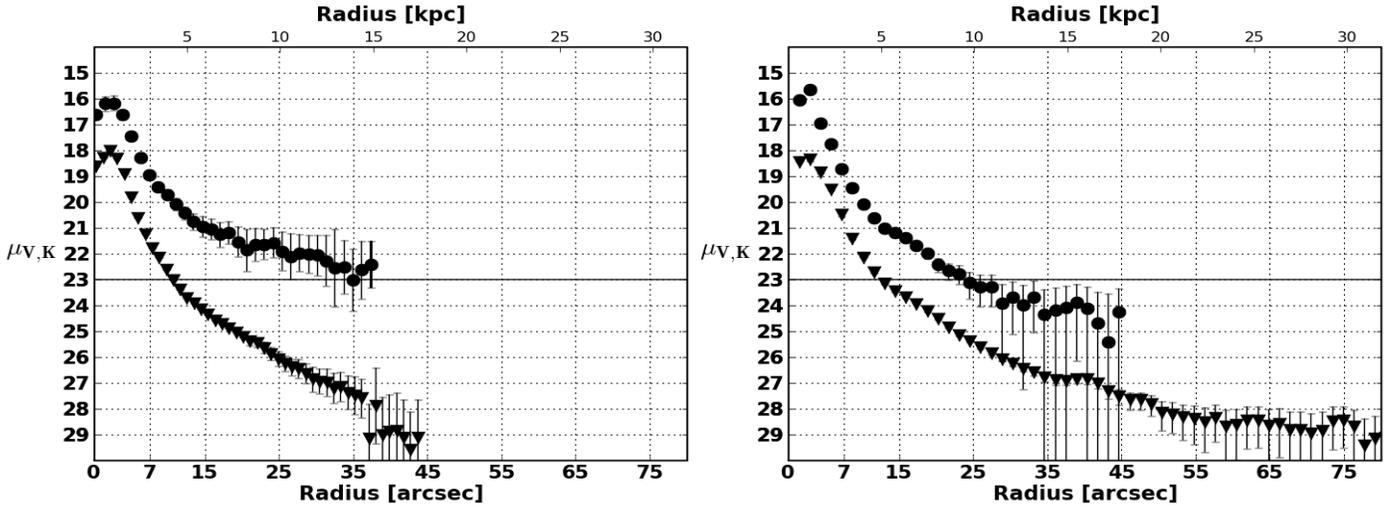}
  \caption{Surface brightness profiles of Haro 11 in V (triangles) and K (circles). \textbf{Left panel:} profiles by \bo~with $e=0.20,~P.A.=120^{\circ}$. The error bars represent the uncertainty in the sky level estimated to be the spread in mean values of a sample of ``empty'' sky regions uniformly distributed over the frames.; \textbf{Right panel:} profiles derived in this work with $e=0.32,~P.A.=103^{\circ}$. The error bars represent the uncertainty in the sky level, $\sigma_{\textrm{sky}}$, together with the uncertainty of the mean flux value of each radius, $\sigma_{\textrm{SDOM}}$. The bumpy feature inside $r<5$\arcsec~ is due to Haro 11's 3 brightest knots~\citep{Cumming_etal_2009}, neither of which is at the center of profile integration. The horizontal line at $\mu=23$ mag\arcsecsq~marks the faintest isophote beyond which the K band profile of an exponential disk systematically deviates from a straight line by more than 0.3 magnitudes. For comparison, the same line is marked in the left panel.}\protect\label{sb}
\end{figure*}

\protect\label{sbsection}
\noindent The surface brightness profiles in the right panel of Fig.~\ref{sb} are obtained by integrating in elliptical rings starting from the center of the galaxy. The faint outskirts of Haro 11 have an almost circular shape in V, however the central starburst is very bright and it is unclear at what radii its contribution to the overall intensity becomes negligible. The shape and center of the underlying host is therefore better estimated from the K band, where the contribution from the starburst quickly drops off with increasing radius. The coordinates of the center and the parameters of the ellipse are chosen by fitting isophotes to the K band image with IRAF ELLIPSE. For robustness, different centering positions, inclinations and position angles (P.A.) have been tested. There is no discernible difference in the radial color profiles obtained from integration with ellipticity and P.A. parameters in the range $e\in(0.20,0.34),~P.A. \in(103^{\circ},120^{\circ}) $, the latter measured from North through East. The range of these values covers all elliptical isophotes returned by IRAF ELLIPSE at surface brightness levels of $\mu_K=21$--$23.5$ mag\arcsecsq, where the signal from the host galaxy dominates the luminosity output. These values are similar to the ones used by \bo~($e=0.20,~P.A.=120^{\circ}$). Once integration starts, the parameters of the elliptic rings and the step size along the major axis are kept constant. The profiles are sampled with a step size of 5 pixels.

\noindent At faint isophotal levels it is the uncertainty in the zero level of the sky background, $\sigma_{sky}$, that largely dominates the error (Fig.~\ref{errplot}). This uncertainty is estimated by masking out all sources, measuring the mean intensity inside square apertures, and calculating the standard deviation of these means. The $\sigma_{sky}$ obtained in this fashion is by definition always numerically greater than the standard deviation of the mean sky level and thus represents a conservative estimate of the sky uncertainty. The apertures are of the same order in size as the area of the smallest elliptic ring inside of the LSB host, measured at $\mu_K\sim21.5$ mag\arcsecsq. Another source of uncertainty included in the error analysis is the uncertainty in the mean flux level of each ring, which is a composite error of the poisson noise and the intrinsic flux scatter across each ring, given by the standard deviation of the mean (SDOM). Instead of isolating the poisson noise in each ring, we use the SDOM in the rings as an upper limit for the poisson noise, because the reduction process invariably destroys the initial poisson distribution of the raw frames. To obtain a more accurate value, one would have to obtain the poisson noise per ring on the raw frames and then propagate those values throughout the reduction process. For the purposes of our analysis, however, the SDOM of each ring on the final reduced frames is an acceptable upper limit. Thus, the combination of the sky uncertainty and the SDOM gives a conservative representation for the uncertainties in our results, summarized in Table~\ref{errtbl}.

\begin{figure}
  \centering
  \vspace{7.5cm}
\includegraphics{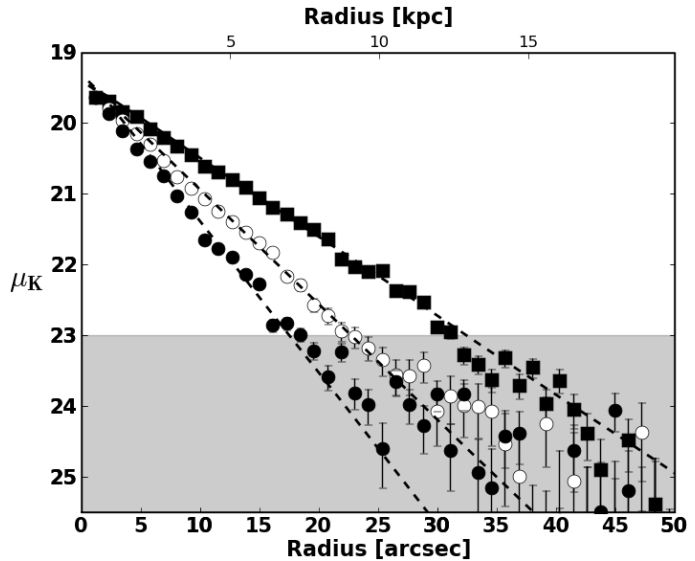}
  \caption[]{Surface brightness profiles of synthetic exponential disks. Filled circles: steep disk with scale length $h_\textrm{r}=2.1$ kpc; open circles: Haro 11-like disk with $h_\textrm{r}=2.7$ kpc; filled squares: flat disk with $h_\textrm{r}=4.0$ kpc. Circles and squares are used for the profiles of the synthetic disks recovered after reduction and calibration. The dashes lines are the analytical profiles of each synthetic disk on an empty frame with no noise. $\mu_{K}=23$ mag\arcsecsq~is the faintest isophote that can be reached by all three exponential disk profiles with negligible random scatter and no systematic effects, see text for justification. This is marked by the gray-shaded area in the figure.
  }\protect\label{fakes}
\end{figure}

\subsection{Setting limits on systematic errors in surface brightness profiles}\protect\label{reliability}

\noindent At low signal-to-noise levels the interpretation of a surface brightness profile is subject to some ambiguity. Observed profile trends could either be real or due to the propagation of systematic errors that have not been accounted for in the reduction process. It is also not always obvious at what surface brightness levels the sky noise starts dominating the behavior of the profile. When pushing surface photometry down to very faint isophotes it is necessary to identify the maximum radial range within which the data are not dominated by any systematic errors that could have been introduced by the reduction process or by the random fluctuations in the sky background. Attempting to ascertain the effects of reduction on the data by analyzing the surface brightness profile of a real physical target is not always practical since the true shape and slope of the profile is often not known in advance. 

\noindent A better way to perform such a test is to insert a synthetic galaxy with an a priori known surface brightness profile in all raw frames, then reduce and calibrate with the same pipeline that was used on the real data. The easiest profile to identify is that of an exponential disk, since it appears as a straight line in a log-lin plot. Analyzing the surface brightness profile of a reduced exponential disk will reveal the presence of any systematic errors, as well as give an indication at what surface brightness levels the sky noise starts dominating the profile. The surface brightness level, at which the reduced synthetic disk profile starts deviating from the expected straight line in a log-lin plot marks the faintest isophote that can be reached with the quality of the data and the accuracy of the pipeline. 

\noindent It is the K band, rather than the V band, that has the noisiest sky and the largest uncertainty in the sky level, so a $V-K$ profile will inevitably be subject to limits imposed solely by the depth and quality of the K band data. We therefore perform the exponential disk test only on the K-band data. For robustness the test is carried out three times, each with different positions and scale lengths of the disks. Fig.~\ref{fakes} shows a plot of the surface brightness profiles of the three reduced exponential disks. The scale lengths are chosen to be steeper, approximately equal to, and flatter than the scale length of the Haro 11 LSB host. The latter is obtained by a least-squares fit to the Haro 11 profile in a range of $\mu_K=21.5$-$23.0$ mag\arcsecsq, which is well away from the central starburst. This gives a Haro 11 scale length of $\approx2.6$ kpc and a y-intercept of $\mu_{0,K}=19.6$ mag\arcsecsq. The central surface brightness is kept constant for all three synthetic disks at $\mu_{0,K}\approx19.5$ mag\arcsecsq. Scale lengths of $2.1$, $2.7$ and $4.0$ kpc are chosen for the steeper, the approximately equal, and the flatter synthetic disks respectively. The shortest and longest scale lengths are chosen to correspond to radial surface brightness profiles that are $\sim1^{\textrm{m}}$ fainter, respectively $\sim1^{\textrm{m}}$ brighter than a profile with a scale length of $2.7$ kpc at a distance of $\textrm{r}=25$ arcsec where the surface brightness is $\mu_K=23$ mag\arcsecsq. This allows the results of the test to be applicable to a range of profile slopes, without the need for great accuracy in the determination of the true scale length of the host galaxy. Indeed, measurements of the Haro 11 scale length from various locations and lengths within $\mu_{K}=20.5$-$23.8$ mag\arcsecsq~give a slightly different value, however the difference is never found to be large enough to displace the profile from within the envelope created by the steep and flat exponential disks included in the test. The K band surface brightness profile for Haro 11 would have to be off by more than $\pm1^{\textrm{m}}$ than what we observe at the radial distance of $\textrm{r}=25$ arcsec, in order for this test to significantly over- or underestimate the K band profile depth that can be reached. 

\noindent Along with the surface brightness profiles of the three reduced exponential disks, Fig.~\ref{fakes} also shows the analytical profiles of each disk on an empty frame. In the absence of any systematic and random uncertainties or in regions where these are insignificant, the reduced exponential disk profiles will exactly follow their analytical counterparts. The steeper and the Haro 11-like disk profiles are consistent with their respective analytical profiles out to $r\sim20.7$ arcsec and $r\sim27.6$ arcsec, respectively, both reaching a depth of $\mu_\textrm{K}\approx23.5$ mag\arcsecsq~with a maximum deviation from their analytical profiles of $\delta\lesssim0.15$ magnitudes at these radii. The flat disk, however, deviates from its analytical counterpart by more than $\delta\approx0.3$ magnitudes beyond $\mu_\textrm{K}\approx23.0$ mag\arcsecsq, $r \gtrsim  32.5$ arcsec. We note that accepting a deviation of $\delta\lesssim0.3$ magnitudes would make this profile consistent with its analytical counterpart down to $\mu_\textrm{K}\approx24.5$ mag\arcsecsq, $r\sim 46.5$ arcsec, however the corresponding maximum deviations of the steep and Haro 11-like profiles are then $\delta\lesssim0.45$ and $\delta\lesssim0.75$ magnitudes, respectively, which we deem to be unacceptably large.

\noindent Analyzing the exact behavior of different exponential disks at extremely low surface brightness levels is, however, beyond the scope of this paper. We are therefore content with using the results of this test to set a conservative limit on the depth of the K band data and we conclude that a surface brightness level of $\mu_{K}=23$ mag\arcsecsq~can clearly be reached for all three profiles with negligible random scatter ($\delta\lesssim0.15$) and no discernible systematic effects. We adopt the limit of $\mu_{K}=23$ mag\arcsecsq~as the faintest isophote that can be reached in the K band with the quality of the Haro 11 data and with the current pipeline. 

\noindent Using a synthetic disk with a Sersic index $n>1$ will not have an effect on the outcome or the interpretation of this test. At the relevant radial distances away from the center all profiles with reasonable values for n are essentially parallel to the exponential disk. Moreover, the exact shape of the analytical profile is irrelevant. It is only important at what radial distance from the center the profile shape can no longer be recovered with acceptable scatter and no systematic effects.

\begin{figure*}
  \centering
  \vspace{7cm}
  \includegraphics{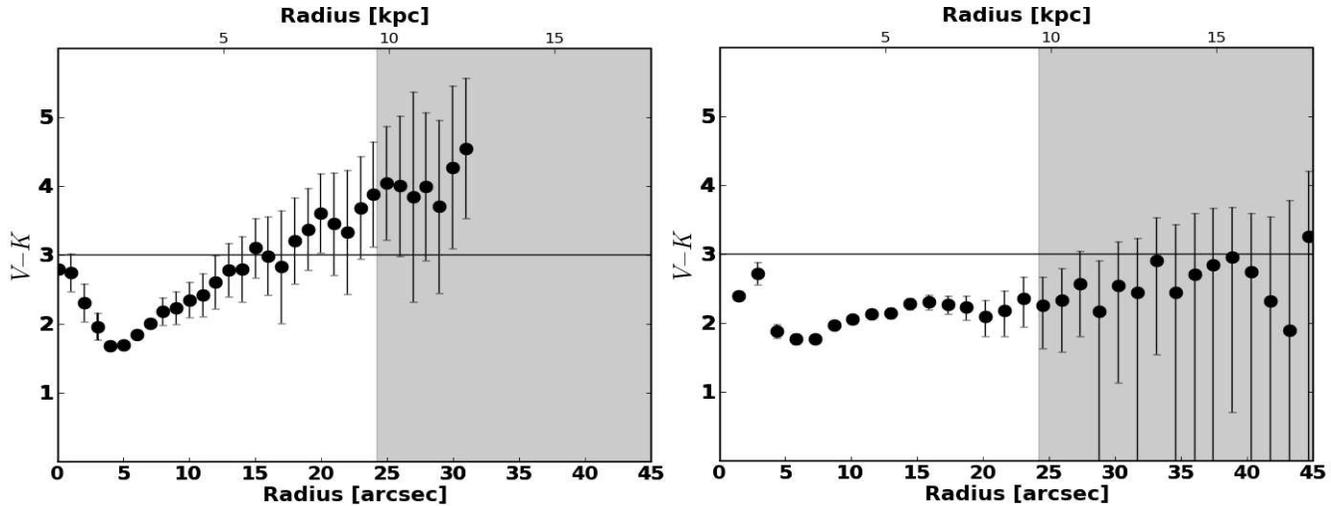}
  \caption{$V-K$ color profiles of Haro 11. \textbf{Left panel:} profile by \bo~ with a step size of $0.95$\arcsec. The error bars represent the uncertainty in the zero level of the sky. No correction for dust reddening is applied. The profile becomes progressively redder with increased radius.; \textbf{Right panel:} profile derived in this work with a step size $1.44$\arcsec. The error bars represent the uncertainty in the sky level, $\sigma_{\textrm{sky}}$, together with the uncertainty of the mean flux value of each radius, $\sigma_{\textrm{SDOM}}$. No correction for dust reddening is applied. The bumpy feature inside $r<5$\arcsec~ is due to Haro 11's 3 brightest knots, neither of which is at the center of profile integration. At a radius of $r\sim14$\arcsec the color stabilizes at $V-K\sim2.3$. The gray-shaded area corresponds to surface brightness magnitudes fainter than $\mu_K=23$~mag\arcsecsq~and is not included in the calculation of the total $V-K$ color. For comparison, the same range is marked in the left panel. }\protect\label{clr}
\end{figure*}

\section[]{Results}\protect\label{results}

\noindent The V and K surface brightness profiles are presented in Fig.~\ref{sb} together with the profiles published by \bo. The V profile is well-behaved and with reasonably small errorbars down to $\mu_V\sim28.5$ mag\arcsecsq. For the K band, however, the synthetic disk tests in $\S$~\ref{reliability} suggest that a K band profile obtained from the current data with the current pipeline is reliable down to $\mu_K\sim23$ mag\arcsecsq. At fainter isophotes the profile is likely to be dominated by systematic errors introduced by the reduction procedure. For the Haro 11 K band profile in Fig.~\ref{sb} $\mu_K=23$ mag\arcsecsq~is reached at a radial distance of $r\sim10$ kpc, or $\sim25$ arcsec, from the center of the galaxy. This radius constitutes the maximum radius that we consider in the consecutive calculation of the total color for the Haro 11 LSB host galaxy.

\noindent Fig.~\ref{clr} shows the $V-K$ radial color profile together with the profile published by \bo. Both profiles are uncorrected for dust extinction and can be directly compared. The two profiles display different behavior. The \bo~profile shows no indication of breaking the trend of becoming progressively redder outwards of $r\ge5$ arcsec, reaching $V-K\simeq3$ at $r\simeq15$ arcsec, while the profile of the new and deeper data reddens from $r\simeq5$ arcsec to $r\simeq15$ arcsec only to flatten out and stabilize at $V-K\simeq2.3$ outwards of $r\ge15$ arcsec. 

\noindent For the calculation of the total $V-K$ halo color we use the radial range $r\simeq5$--$10$ kpc. The K band profile changes slope at $r\simeq5$ kpc and seems consistent with a constant slope out to $r\simeq13$ kpc. This suggests that the contribution from the central starburst in the K band is negligible already at $r\simeq5$ kpc and beyond this radius the profile is instead dominated by the LSB host. We therefore take $r_0\simeq5$ kpc as the starting radius for the calculation of the total color. For the maximum radius we take the limit of $r_\textrm{max}\simeq10$ kpc, as previously justified in $\S$~\ref{reliability}. Over this radial range, we obtain a total color of $V-K=2.3\pm0.2$. This is $2.3\sigma$ away from the \bo~value of $V-K=4.2\pm0.8$.

\noindent It should be noted that the total $V-K$~color reported in \bo, $V-K=4.2\pm0.8$, is measured out to a distance of $r\simeq12.3$ kpc. This is outside of the maximum radius limit we impose on our data, $r_\textrm{max}\simeq10$ kpc. For direct comparison of the two total color measurements, we also measure the total color out to a radius of $r\simeq12.3$ kpc as in \bo, and obtain the slightly higher value for the total color of $V-K=2.4\pm0.5$, which is $2\sigma$ away from the total color of \bo. We cannot justify the use of data points out to such large radii in the total $V-K$ color because our exponential disk test with a Haro 11-like disk shows a systematic brightening of the K band surface brightness profile beyond $r\gtrsim11$ kpc (Fig.~\ref{fakes}). In the absence of a similar systematic effect in the V band at the same radii, the brightening of the K band profile will result in a spurious reddening of the $V-K$ radial profile.

\noindent We further note that a simple least-squares fit of the Haro 11 profile in the range of $\mu_K=21.5$-$23.0$ mag\arcsecsq~gives a scale length of the host population of $\simeq2.6$ kpc in the K band with an extrapolated central surface brightness of $\mu_{0,K}\simeq19.6$ mag\arcsecsq.

\begin{figure}
\centering
\includegraphics[width=8cm]{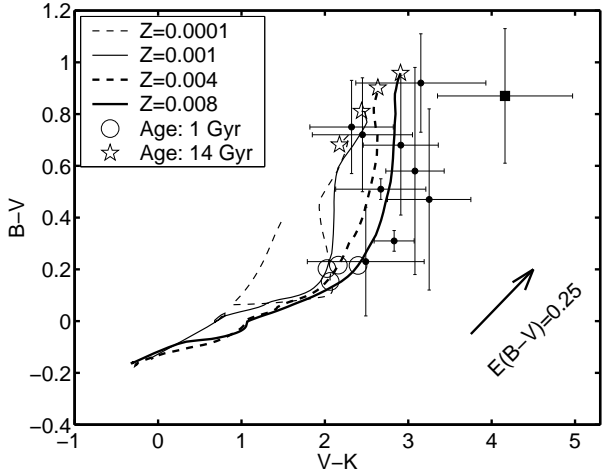}
\caption{\protect\cite{Marigo_etal_2008} stellar evolutionary tracks for different metallicities with halo color data for all 10 BCGs of the \bo~and~\citet{Bergvall_etal_2005} sample. The old color $V-K=4.2\pm0.8$ for Haro 11 is marked with a filled square. The tracks clearly cannot reproduce this data point.}
\protect\label{SEM_old}
\end{figure}
\begin{figure}
\centering
\includegraphics[width=8cm]{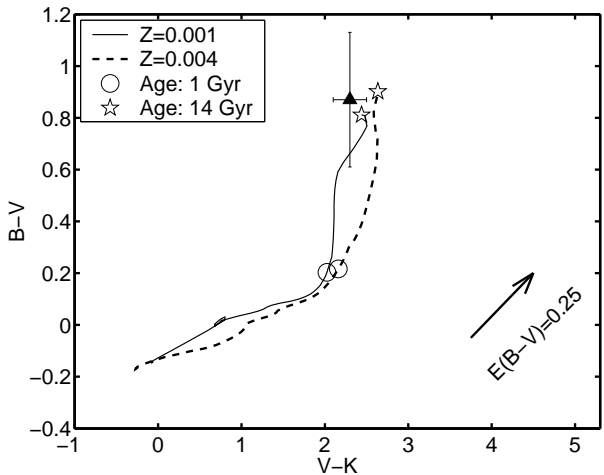}
\caption{\protect\cite{Marigo_etal_2008} stellar evolutionary tracks for $Z=0.001$ and $Z=0.004$ with the new Haro 11 halo color of $V-K=2.3\pm0.2$ marked with a filled triangle. The data point is consistent with an old stellar population of low metallicity, $Z=0.001$.}
\protect\label{SEM_new}
\end{figure}
\section[]{Discussion}\protect\label{discuss}
\noindent In this section we examine how our results compare with previously published measurements and comment on the implications of the new Haro 11 host color for the underlying stellar population.

\subsection[]{Optical/near-IR color interpretation}
\noindent The previous color measurement of the LSB host of Haro 11, $V-K=4.2\pm0.8$, was part of a larger deep optical/near-IR photometric study of four luminous metal-poor BCGs~\citep{Bergvall_Ostlin_2002} and an additional six metal-poor BCGs~\citep{Bergvall_etal_2005}. The colors for all galaxies in the sample were found to be consistently too red to reconcile with a normal stellar population of low metallicity. This has lead to somewhat controversial scenarios for the nature of the underlying stellar population~\citep{Bergvall_Ostlin_2002,Zackrisson_etal_2006,Zackrisson_etal_2009b}. The measured colors could be modeled with a low-metallicity ($Z=0.001$) stellar population with a very bottom-heavy IMF ($dN/dM\propto M^{-\alpha}$ with $\alpha=4.50$) and some dust reddening.~\citet{Zackrisson_etal_2006} found that in addition to these BCGs, the red colors of the faint outskirts of the stacked edge-on disks of~\citet{Zibetti_etal_2004} could also be explained by such an IMF. This solution also works for the~\citeauthor*{BergvallZackrissonCaldwell2009}~\citeyearpar{BergvallZackrissonCaldwell2009} detection, since the halo colors of their SDSS stack are similar to those of ~\citet{Zibetti_etal_2004}. If such red halos are found to be a common feature of galaxies of very different morphological types, this could substantially contribute to the baryonic dark matter content of the Universe, supporting the idea that a non-negligible fraction of the missing baryons is locked up in faint low-mass stars. These low-mass stars cannot be smoothly distributed over the halo, but would instead need to be concentrated in star clusters, since a smooth halo component with a bottom-heavy IMF is ruled out by Milky Way star count data~\citep{Zackrisson_Flynn_2008}.

\noindent After the improved implementation of the thermally pulsing AGB phase in stellar evolutionary models~\citep{Marigo_etal_2008} all BCG galaxies in the sample, except Haro 11, could be reconciled with a stellar population with a Salpeter IMF of an intermediate metallicity of $Z=0.004$--$0.008$, or $20$--$40\%$ of $Z_{\odot}\sim0.02$~\citep{Zackrisson_etal_2009b}, with the caveat that these inferred stellar metallicities of the LSB hosts are higher than the observed gas metallicities of the central starburst regions (\bo, $Z=0.002$, $10\%$ of $Z_{\odot}$). Haro 11 had the reddest optical/near-IR color from the sample, $V-K=4.2\pm0.8$, which could not be explained with a Salpeter IMF. Figure~\ref{SEM_old} shows the~\cite{Marigo_etal_2008} stellar evolutionary tracks for all galaxies in the sample for a Salpeter IMF. Haro 11 is clearly an outlier and cannot be reconciled with any reasonable metallicity. The only model that provides a reasonable fit for $V-K=4.2\pm0.8$ is the bottom-heavy IMF model, albeit with somewhat high metallicity~\citep{Zackrisson_etal_2006}.

\noindent Analysis of the deep new data presented in this paper suggests that the color of the host galaxy of Haro 11 is instead $V-K=2.3\pm0.2$. This revised color is shown in Fig.~\ref{SEM_new} with the~\cite{Marigo_etal_2008} stellar evolutionary tracks for a Salpeter IMF for low and intermediate metallicity. We do not have new B band data for Haro 11 which would allow us to reduce the uncertainty in the $B-V$ color, therefore we cannot give a better age estimate for the LSB host. However, we have reduced the uncertainty in the $V-K$ color by a factor of 4 and from Fig.~\ref{SEM_new} we can draw the conclusion that the host galaxy is consistent with a metal-poor, very old, but normal IMF stellar population. 

\subsection[]{Comparison with previous observations}
\noindent The discrepancy between our and the \bo~color measurement is $\sim2\sigma$. The source of this discrepancy is most likely the improved quality of CCDs and NIR arrays, as well as the larger detector sizes of the new observations. The most dramatic change in data quality is seen in the K band (Figs.~\ref{VK},~\ref{Kband}). Because of the higher sky brightness and the lower atmospheric transmission in the K band, ground-based V band observations are usually $\sim5$ (Vega) magnitudes deeper than ground-based K band data. The behavior of a $V-K$ color profile at very faint isophotes is thus always subject to limitations imposed by the depth of the K band data. Given favorable observing conditions and assuming the reduction is carried out correctly, this in turn would depend on the detector quality, the total exposure time, and the specific morphology of the target galaxy. Since the morphology of Haro 11 has not changed in the last decade, we examine the effects of the increase in exposure time. For the K band data the effective exposure time (taking telescope aperture into account) is now 2.4 times longer than the previously published data. Comparison of the two K surface brightness profiles in Fig.~\ref{sb} shows that the added detector integration has increased the radial range within which the K profile is accurate to more than $\pm0.5$ mag by $\approx2$ kpc. To test whether the longer exposure time also accounts for the different slope of the profiles at large radii, we compared a K band surface brightness profile obtained with $100\%$ of all new K band frames to a profile obtained with only $40\%$ of the frames, equivalent to an effective exposure time of $25$ minutes (25 minutes on the ESO NTT correspond to $66$ minutes on the ESO 2.2m). We conclude that there is no indication that a shallower profile would be systematically brighter at smaller radii than a deeper profile. The effect is simply an increase in the random noise at smaller radii but the overall shape and slope of the profile at faint isophotes show no significant change. 
On the other hand, an insufficient sky subtraction would cause a brightening in the surface brightness profile at large radii from the center of the galaxy. To test this, we have simulated a sky subtraction problem by adding a constant sky residual to each data point of the ESO NTT SOFI profile seen in the right panel of Fig.~\ref{sb}. We find that in the framework of the ESO 2.2m IRAC2 data, a sky residual on the order of $\sim1.4\sigma$ is needed to account for the observed brightening of the ESO 2.2m IRAC2 surface brightness profile seen in the left panel of Fig.~\ref{sb}. We therefore conclude that the shape of the ESO 2.2m IRAC2 K band profile is most likely due to the lingering presence of a sky residual of $\sim1.4\sigma$. Over the same radial range, this sky subtraction problem has been avoided in the ESO NTT SOFI profile mainly due to the improved observing techniques, the better quality of the NIR array, improved sky region estimation techniques and the use of a larger light-gathering area. Specifically, the effective field of view of the final stacked image has increased by a factor of 2.3, thus providing a comparatively larger area for sky estimation, well away from the faintest outskirts of Haro 11 visible in the K band.

\noindent The work in this paper indicates the need for a possible revision of the color measurements for the 10 BCGs in the \bo~and~\citet{Bergvall_etal_2005} sample. Currently underway is the reduction of a sample of nearly 40 BCGs in the optical and NIR, which also includes new and deeper data for 5 of the 6 BCGs in the \bo~sample. 

\section[]{Summary}\protect\label{summary}
\noindent We have presented deep new optical/near-IR data of the BCG Haro 11 and performed surface photometry on the faint outskirts of the underlying LSB host galaxy. We have obtained a total color of $V-K=2.3\pm0.2$ for the host over the range  $r=5$--$10$ kpc. This is $2.3\sigma$ off from the previously published measurement of $V-K=4.2\pm0.8$, which was measured out to $r=12.3$ kpc. Increasing the radial range of the total color measurement out to $r=12.3$ kpc gives $V-K=2.4\pm0.5$ which has a $\sim2\sigma$ discrepancy to previous results. The new color places Haro 11 on the blue end of the $V-K$ range occupied by the other BCGs in the sample and reconciles the LSB host of Haro 11 with a metal-poor ($Z=0.001$) stellar population with a standard Salpeter IMF. Hence, the metallicity indicated by the color is consistent with that measured from emission-line ratios, and an anomalous stellar population is no longer required to explain the properties of the Haro 11 host galaxy.

\section*{Acknowledgments}

\noindent The near-IR observations in this work are made with the ESO NTT telescope at the La Silla Observatory under program ID 075.B-0220. The optical observations in this work are made with the Nordic Optical Telescope, operated on the island of La Palma jointly by Denmark, Finland, Iceland, Norway, and Sweden, in the Spanish Observatorio del Roque de los Muchachos of the Instituto de Astrofisica de Canarias.

\noindent The authors would like to thank the anonymous referee for thought-provoking comments, and the editor Anna Evripidou for her expeditious help. EZ acknowledges a research grant from the Swedish Royal Academy of Sciences. G\"O and NB acknowledge support from the Swedish Research Council. G\"O is a Royal Swedish Academy of Sciences Research Fellow supported by a grant from the Knut and Alice Wallenberg Foundation.

\bibliographystyle{mn2e}
\bibliography{redhaloHaro11}

\begin{thebibliography}{}

\bibitem[\protect\citeauthoryear{{Amor{\'{\i}}n}, {Aguerri},
  {Mu{\~n}oz-Tu{\~n}{\'o}n} \& {Cair{\'o}s}}{{Amor{\'{\i}}n}
  et~al.}{2009}]{Amorin_etal_2009}
{Amor{\'{\i}}n} R.,  {Aguerri} J.~A.~L.,  {Mu{\~n}oz-Tu{\~n}{\'o}n} C.,
  {Cair{\'o}s} L.~M.,  2009, ArXiv e-prints, arXiv:0903.2861

\bibitem[\protect\citeauthoryear{{Amor{\'{\i}}n}, {Mu{\~n}oz-Tu{\~n}{\'o}n},
  {Aguerri}, {Cair{\'o}s} \& {Caon}}{{Amor{\'{\i}}n}
  et~al.}{2007}]{Amorin_etal_2007}
{Amor{\'{\i}}n} R.~O.,  {Mu{\~n}oz-Tu{\~n}{\'o}n} C.,  {Aguerri} J.~A.~L.,
  {Cair{\'o}s} L.~M.,    {Caon} N.,  2007, \aap, 467, 541

\bibitem[\protect\citeauthoryear{{Bergvall}, {Marquart}, {Persson},
  {Zackrisson} \& {{\"O}stlin}}{{Bergvall} et~al.}{2005}]{Bergvall_etal_2005}
{Bergvall} N.,  {Marquart} T.,  {Persson} C.,  {Zackrisson} E.,    {{\"O}stlin}
  G.,  2005, in {Renzini} A.,  {Bender} R.,  eds, Multiwavelength Mapping of
  Galaxy Formation and Evolution {Strange Hosts of Blue Compact Galaxies}.
pp 355--+

\bibitem[\protect\citeauthoryear{{Bergvall} \& {{\"O}stlin}}{{Bergvall} \&
  {{\"O}stlin}}{2002}]{Bergvall_Ostlin_2002}
{Bergvall} N.,  {{\"O}stlin} G.,  2002, \aap, 390, 891

\bibitem[\protect\citeauthoryear{{Bergvall}, {Zackrisson} \&
  {Caldwell}}{{Bergvall} et~al.}{2009}]{BergvallZackrissonCaldwell2009}
{Bergvall} N.,  {Zackrisson} E.,    {Caldwell} B.,  2009, ArXiv e-prints,
  arXiv:0909.4296

\bibitem[\protect\citeauthoryear{{Cair{\'o}s}, {V{\'{\i}}lchez}, {Gonz{\'a}lez
  P{\'e}rez}, {Iglesias-P{\'a}ramo} \& {Caon}}{{Cair{\'o}s}
  et~al.}{2001}]{Cairos_etal1_2001}
{Cair{\'o}s} L.~M.,  {V{\'{\i}}lchez} J.~M.,  {Gonz{\'a}lez P{\'e}rez} J.~N.,
  {Iglesias-P{\'a}ramo} J.,    {Caon} N.,  2001, \apjs, 133, 321

\bibitem[\protect\citeauthoryear{{Cumming}, {{\"O}stlin}, {Marquart}, {Fathi},
  {Bergvall} \& {Adamo}}{{Cumming} et~al.}{2009}]{Cumming_etal_2009}
{Cumming} R.~J.,  {{\"O}stlin} G.,  {Marquart} T.,  {Fathi} K.,  {Bergvall} N.,
     {Adamo} A.,  2009, ArXiv e-prints, arXiv:0901.2869

\bibitem[\protect\citeauthoryear{{de Jong}}{{de Jong}}{2008}]{deJong_2008}
{de Jong} R.~S.,  2008, \mnras, 388, 1521

\bibitem[\protect\citeauthoryear{{Doublier}, {Comte}, {Petrosian}, {Surace} \&
  {Turatto}}{{Doublier} et~al.}{1997}]{Doublier_etal_1997}
{Doublier} V.,  {Comte} G.,  {Petrosian} A.,  {Surace} C.,    {Turatto} M.,
  1997, \aaps, 124, 405

\bibitem[\protect\citeauthoryear{{Hook}, {Maisala}, {Oittinen}, {Ullgren},
  {Vasko}, {Savolainen}, {Lindroos}, {Anttila}, {Solin}, {M{\o}ller}, {Banse}
  \& {Peron}}{{Hook} et~al.}{2006}]{Hook_etal_2006}
{Hook} R.~N.,  {Maisala} S.,  {Oittinen} T.,  {Ullgren} M.,  {Vasko} K.,
  {Savolainen} V.,  {Lindroos} J.,  {Anttila} M.,  {Solin} O.,  {M{\o}ller}
  P.~M.,  {Banse} K.,    {Peron} M.,  2006, in {Gabriel} C.,  {Arviset} C.,
  {Ponz} D.,   {Enrique} S.,  eds, Astronomical Data Analysis Software and
  Systems XV Vol.~351 of Astronomical Society of the Pacific Conference Series,
  {PyMidas--A Python Interface to ESO-MIDAS}.
pp 343--+

\bibitem[\protect\citeauthoryear{{Kunth} \& {{\"O}stlin}}{{Kunth} \&
  {{\"O}stlin}}{2000}]{Kunth_Ostlin_2000}
{Kunth} D.,  {{\"O}stlin} G.,  2000, \aapr, 10, 1

\bibitem[\protect\citeauthoryear{{Loose} \& {Thuan}}{{Loose} \&
  {Thuan}}{1986}]{Loose_Thuan_1986}
{Loose} H.-H.,  {Thuan} T.~X.,  1986, \apj, 309, 59

\bibitem[\protect\citeauthoryear{{Marigo}, {Girardi}, {Bressan}, {Groenewegen},
  {Silva} \& {Granato}}{{Marigo} et~al.}{2008}]{Marigo_etal_2008}
{Marigo} P.,  {Girardi} L.,  {Bressan} A.,  {Groenewegen} M.~A.~T.,  {Silva}
  L.,    {Granato} G.~L.,  2008, \aap, 482, 883

\bibitem[\protect\citeauthoryear{{Noeske}, {Papaderos}, {Cair{\'o}s} \&
  {Fricke}}{{Noeske} et~al.}{2005}]{Noeske_etal_2005}
{Noeske} K.~G.,  {Papaderos} P.,  {Cair{\'o}s} L.~M.,    {Fricke} K.~J.,  2005,
  \aap, 429, 115

\bibitem[\protect\citeauthoryear{{Papaderos}, {Loose}, {Thuan} \&
  {Fricke}}{{Papaderos} et~al.}{1996}]{Papaderos_etal1_1996}
{Papaderos} P.,  {Loose} H.-H.,  {Thuan} T.~X.,    {Fricke} K.~J.,  1996,
  \aaps, 120, 207

\bibitem[\protect\citeauthoryear{{Zackrisson}, {Bergvall}, {{\"O}stlin},
  {Micheva} \& {Leksell}}{{Zackrisson} et~al.}{2006}]{Zackrisson_etal_2006}
{Zackrisson} E.,  {Bergvall} N.,  {{\"O}stlin} G.,  {Micheva} G.,    {Leksell}
  M.,  2006, \apj, 650, 812

\bibitem[\protect\citeauthoryear{{Zackrisson} \& {Flynn}}{{Zackrisson} \&
  {Flynn}}{2008}]{Zackrisson_Flynn_2008}
{Zackrisson} E.,  {Flynn} C.,  2008, \apj, 687, 242

\bibitem[\protect\citeauthoryear{{Zackrisson}, {Micheva}, {Bergvall} \&
  {\"Ostlin}}{{Zackrisson} et~al.}{2009}]{Zackrisson_etal_2009b}
{Zackrisson} E.,  {Micheva} G.,  {Bergvall} N.,    {\"Ostlin} G.,  2009, ArXiv
  e-prints, arXiv:0902.4695

\bibitem[\protect\citeauthoryear{{Zackrisson}, {Micheva} \&
  {{\"O}stlin}}{{Zackrisson} et~al.}{2009}]{Zackrisson_etal_2009a}
{Zackrisson} E.,  {Micheva} G.,    {{\"O}stlin} G.,  2009, \mnras, 397, 2057

\bibitem[\protect\citeauthoryear{{Zibetti} \& {Ferguson}}{{Zibetti} \&
  {Ferguson}}{2004}]{Zibetti_Ferguson_2004}
{Zibetti} S.,  {Ferguson} A.~M.~N.,  2004, \mnras, 352, L6

\bibitem[\protect\citeauthoryear{{Zibetti}, {White} \& {Brinkmann}}{{Zibetti}
  et~al.}{2004}]{Zibetti_etal_2004}
{Zibetti} S.,  {White} S.~D.~M.,    {Brinkmann} J.,  2004, \mnras, 347, 556

\end{thebibliography}

\bsp

\label{lastpage}

\end{document}